\documentclass[english,aps,preprint]{revtex4}
\usepackage[T1]{fontenc}
\usepackage[latin1]{inputenc}
\usepackage{graphicx}

\makeatletter

\usepackage{babel}
\makeatother
\begin{document}

\title{Lattice formulation of the Fokker-Planck equation}

\author{Simone Melchionna}

\affiliation{INFM-SOFT, Department of Physics, University of Rome {}``La Sapienza'',
P.le A. Moro 2, 00185 Rome, Italy}

\author{Sauro Succi}

\affiliation{Istituto per le Applicazioni del Calcolo ``A. Picone'', CNR, Viale
del Policlinico 137, 00166, Roma, Italy}

\author{Jean-Pierre Hansen}

\affiliation{Department of Chemistry, Lensfield Road, Cambridge CB2 1EW, United
Kindgom}

\begin{abstract}
A lattice version of the Fokker-Planck equation (FPE), accounting
for dissipative interactions, not resolved on the molecular scale,
is introduced. The lattice FPE is applied to the study of electrorheological
transport of a one-dimensional charged fluid, and found to yield quantitative
agreement with a recent analytical solution. Future extensions, including
inelastic ion-ion collisions, are also outlined.
\end{abstract}
\maketitle

\section{introduction}

Over the last decade discrete lattice versions of kinetic equations,
most notably the Lattice-Boltzmann (LB) method, have undergone burgeoning
progress for the simulation of large scale hydrodynamic flows \cite{benzi-succi-vergassola,wolfe-gladrow,succi-book,chen-doolen}
and of the dynamics of colloidal suspensions \cite{ladd-verberg,cates}.
The LB method was developed in response to major flaws of its predecessor,
the lattice gas cellular automaton \cite{frisch-hasslacher-pomeau,eu},
following the canonical route of classical Statistical Mechanics.
A few years later the formal connection of the LB method with continuum
kinetic theory was also elucidated \cite{toadd}.

One of the major appeals of the LB method is its flexibility, which
allows to accomodate a host of complex physical effects, including
boundary conditions at interfaces, intermolecular forces and even
chemical reactions, through efficient and elegant discretizations
of the force term in the kinetic Vlasov-Boltzmann equation. One of
the limitations of the LB method is that, since it generates the time
evolution of the one particle distribution function, fluctuations
are not accounted for. {}``Brownian noise'' becomes increasingly
important as one explores flows on ever smaller scales, as in colloidal
systems, or in narrow pores (microfluidics). The problem has been
addressed on the colloidal scales, by incorporating a random (Brownian)
force component in the momentum flux stress tensor \cite{ladd,ladd-verberg},
or at the level of the discrete LB equation itself \cite{cates}.

On the other hand, all LB implementations to date are based on the
relaxation form of the collision operator, either in scalar \cite{bgk}
or tensorial form \cite{toadd2}. In this paper we show that the LB
methodology can be easily extended to deal with small scale processes
involving a Brownian component, by introducing a lattice version of
the Fokker-Planck (FP) collision operator \cite{vankampen,risken}.
In full analogy with the LB equation, the present lattice Fokker-Planck
equation builds upon an optimized form of importance sampling of velocity
space which, at variance with most numerical grid methods \cite{risken,zhang-wei-kouri-hoffman,fokguotang},
permits to solve the Fokker-Planck equation near local-equilibrium
in full single-particle phase space. The FP operator accounts for
frictional dissipation due, e.g., to solute-solvent interactions or
inelastic collisions with obstacles and confining surfaces, which
need not be resolved on a molecular scale. We have, in particular,
in mind electrorheological transport of ions confined in swollen clays,
between membranes, or through water-filled nanopores (see e.g. \cite{melchionna-succi}),
like ion channels through membranes \cite{Hille}. The FP equation
has recently been applied to a detailed analysis of stationary ion
currents through one-dimensional pores of finite or infinite length
\cite{piasecki-allen-hansen}, and analytic results were obtained
in the case of independent (non-interacting) ions. In particular the
ion current was shown to saturate with increasing applied field, and
this result will serve as a benchmark for the numerical calculations
presented later in this paper.

The paper is organized as follows. In section 2 we discuss the respective
roles of the FP and Bhatnagar-Gross-Krook (BGK) collision operators
in modelling transport and flows of solutes in implicit solvent and
confined geometries. A one-dimensional kinetic model for ion transport
through narrow pores is presented in Section 3. The lattice Fokker-Planck
(LFP) equation for this problem is derived in Section 4. Numerical
results are presented in Section 5 and compared to the predictions
of ref. \cite{piasecki-allen-hansen}, while concluding remarks are
made in Section 6. Stability analysis of the LFPE is outlined in the
appendix.

\section{Fokker-Planck versus BGK collision operators}

Kinetic equations for the time evolution of the distribution function
$f({\bf r},{\bf v};t)$ of a given species of particles conventionally
involve a free-flow term (left-hand side) and a collision term (right-hand
side), i.e.\begin{equation}
(\partial_{t}+v_{\alpha}\partial_{x_{\alpha}}+a_{\alpha}\partial_{v_{\alpha}})f({\bf r},{\bf v};t)=C[f({\bf r},{\bf v};t)]\end{equation}
where $x_{\alpha}$ and $v_{\alpha}$ ($1\leq\alpha\leq d$) are the
cartesian components of the d-dimensional position and velocity vectors
${\bf r}$ and ${\bf v}$, $\partial_{x_{\alpha}}$ and $\partial_{v_{\alpha}}$
are the components of the corresponding gradient operators, while
$a_{\alpha}=F_{\alpha}/m$ are the components of the acceleration
due to an externally applied, or a self-consistent force field ${\bf F}$
($m$ is the particle mass). The Einstein convention of summation
over repeated indices $\alpha$ is assumed. On the right hand side
$C$ denotes a collision operator (yet to be specified), acting on
the distribution function $f$.

If one is interested in flows involving two (or more) species, e.g.
a solvent and a solute, one may adopt one of two strategies: 

a) one may treat the two species (say $a$ and $b$) on the same footing,
by introducing two distribution functions $f_{i}({\bf r},{\bf v};t)$
($i=a$ or $b$) , associated with the two species, the time evolution
of which is governed by two coupled kinetic equations. In the perspective
of a discrete lattice formulation of the LB type, the simplest collision
operators $C_{ij}$($i,j=a$ or $b$) acting on the distribution functions
are of the BGK form, involving several relaxation times \cite{luo-girimaji}.

b) An alternative route, which is particularly appropriate when the
solutes are colloidal particles, and which will be adopted here, is
to assume a separation of time scales, and use an implicit solvent
description of the Langevin form, involving frictional and random
forces. The corresponding collision operator is the familiar Fokker-Planck
operator \cite{vankampen,risken} which may also account for inelastic
collisions of the solute particles with confining surfaces. In such
an effective one-component description, the \emph{dissipative} solute/solvent
and solute/wall couplings are described by a FP collision operator,
while \emph{conservative} solute/solute collisions are described by
a BGK operator. Explicitly, in the presence of the electric field,\begin{equation}
(\partial_{t}+v_{\alpha}\partial_{x_{\alpha}}+a_{\alpha}\partial_{v_{\alpha}})f({\bf r},{\bf v};t)=C^{FP}[f({\bf r},{\bf v};t)]+C^{BGK}[f({\bf r},{\bf v};t)]\label{eq:model}\end{equation}
where\begin{equation}
C^{FP}[f]=\partial_{v_{\alpha}}(R_{\alpha}+D\partial_{v_{\alpha}})f\label{eq:FPoperator}\end{equation}
\begin{equation}
C^{BGK}[f]=-\omega(f-f_{BGK}^{eq})\label{eq:BGKoperator}\end{equation}
In eq. (\ref{eq:FPoperator}) $R_{\alpha}$ are the components of
the drag force, which will be taken of the familiar form ${\bf R}=\gamma{\bf v}$,
where $\gamma$ is a constant friction coefficient, $D$ characterizes
diffusion in velocity space, and is related to $\gamma$ by $D=\gamma v_{T}^{2}$,
where $v_{T}=\sqrt{k_{B}T/m}$ is the thermal velocity and $\omega=1/\tau$
is the solute-solute collision frequency. 

The BGK collision operator entails a relaxation of the distribution
function to the usual local Maxwellian equilibrium,

\begin{equation}
f_{BGK}^{eq}({\bf r},{\bf v};t)=\frac{n({\bf r};t)}{(2\pi v_{T}^{2})^{3/2}}e^{-({\bf v}-{\bf u})^{2}/2v_{T}^{2}}\end{equation}
where $n({\bf r};t)$ is the local density, ${\bf u}={\bf u}({\bf r};t)$
is the local average (or flow) velocity, while the thermal velocity
$v_{T}$ may depend on position and time, via the local temperature
$T({\bf r};t)$.

The equilibrium resulting from the combined action of Fokker-Planck
collisions and the electric field is given by a global shifted Maxwellian:
\begin{equation}
f_{FP,E}^{eq}({\bf r},{\bf v};t)=\frac{n({\bf r};t)}{(2\pi v_{T}^{2})^{3/2}}e^{-(\mathbf{v}-\mathbf{u}_{E})^{2}/2v_{T}^{2}}\end{equation}
where $\mathbf{u}_{E}\equiv\frac{q\mathbf{E}}{m\gamma}$ is the drift
speed associated with the electric field. By definition, the composite
local equilibrium resulting from the competition of the two operators
in the presence of an electric field satisfies the condition\begin{equation}
C^{FP}[f^{eq}]+C^{BGK}[f^{eq}]-a_{\alpha}\partial_{v_{\alpha}}f^{eq}=0\end{equation}
This may be rewritten as\begin{equation}
\mathcal{C}^{FP,E}[f^{eq}]=\frac{\omega}{\gamma}(f^{eq}-f_{BGK}^{eq})\label{eq:FEQ}\end{equation}
where $\mathcal{C}^{FP,E}=\partial_{v_{\alpha}}(v_{\alpha}+v_{T}^{2}\partial_{v_{\alpha}})-(a_{\alpha}/\gamma)\partial_{v_{\alpha}}$is
the Fokker-Planck operator including the electric field.

From the expression (\ref{eq:FEQ}), it is clear that in the strongly
dissipative regime $\omega/\gamma\rightarrow0$, $f^{eq}\rightarrow f_{FP,E}^{eq}$,
while the leading order correction is proportional to the difference
$f_{FP,E}^{eq}-f_{BGK}^{eq}$, i.e. to the deviation of the local
flow field $\mathbf{u}$ from the drift speed $\mathbf{u}_{E}$. This
is consistent with the fact that when $\mathbf{u}=\mathbf{u}_{E}$,
a condition which is reached at steady-state, the FP and BGK equilibria
coincide.

The interplay between the FP and BGK collision operators spawns a
rich variety of physical effects, which will be the object of future
investigations. In the sequel, however, we shall confine our attention
to the methodological aspects related to the lattice formulation of
the Fokker-Planck equation. For illustration purposes, and the sake
of simplicity, we henceforth restrict the discussion to the case $d=1$
(one spatial dimension) of the FP equation and to its validation by
comparison with recent exact results.

\section{A one-dimensional kinetic model for transport through pores}

In this Section the implicit solvent kinetic model introduced in the
previous Section is applied to the important problem of single-file
ion transport through a water-filled pore connecting two reservoirs,
under the action of an applied electric field or ion-concentration
gradient. The one-dimensional version of the kinetic model (\ref{eq:model})-(\ref{eq:BGKoperator})
provides a crude representation of ion permation of ion channels through
membranes separating intra and extra-cellular compartments \cite{Hille,piasecki-allen-hansen}.
Ion permeation of such channels has been examined by numerous Molecular
Dynamics (MD) or Brownian Dynamics (BD) simulations of realistic or
semi-realistic quasi-cylindrical models (for a review, see \cite{tieleman-biggins-smith-samson})
or by numerical solutions of the Poisson-Nernst-Planck equations \cite{chen-eisenberg},
but the present study is inspired by the recent kinetic modelling
of ref. \cite{piasecki-allen-hansen}.

The action of the confining, quasi-cylindrical pore is crudely represented
by restricting ion motion to one dimension and by a contribution to
the frictional force $-\gamma v$.

The one-dimensional version of the kinetic equation (\ref{eq:model})-(\ref{eq:BGKoperator})
may be cast in the form\begin{equation}
(\partial_{t}+v\partial_{x})f=C^{FP}[f]+C^{BGK}[f]-a\partial_{v}f=\partial_{v}[\gamma(vf+v_{T}^{2}\partial_{v}f)]-\omega(f-f_{BGK}^{eq})-a\partial_{v}f\label{eq:1dmodel}\end{equation}
where $f=f(x,v;t)$ and $a=qE/m$, $E$ being the applied electric
field and $q$ the charge of the ions; in practice one is mostly interested
in mono or divalent cations ($q=+e$ or $+2e$). The last two terms
on the r.h.s. of eq. (\ref{eq:1dmodel}) may be regrouped into a single
FP-like term \begin{equation}
C^{FP}[f]-a\partial_{v}f=\partial_{v}\gamma[(v-u_{E})f+v_{T}^{2}\partial_{v}f]\label{eq:1dFPoperator}\end{equation}
where $u_{E}=qE/m\gamma=a/\gamma$ is the ion drift velocity in response
to the applied field. 

The local equilibrium solution of the kinetic equation (\ref{eq:1dmodel})
is hence\begin{equation}
f_{BGK}^{eq}(x,v;t)=\frac{n(x;t)}{(2\pi v_{T}^{2})^{1/2}}e^{-[v-u(x;t)]^{2}/2v_{T}^{2}}\end{equation}
The zeroth, first and second moments of the distribution are the local
density $n$, current $J$ and pressure (or momentum flux) $P$ per
unit mass\begin{equation}
n(x;t)=\int_{-\infty}^{\infty}f(x,v;t)dv\label{eq:density}\end{equation}
\begin{equation}
J(x;t)=\int_{-\infty}^{\infty}vf(x,v;t)dv\equiv n(x;t)u(x;t)\label{eq:current}\end{equation}
\begin{equation}
P(x;t)=\int_{-\infty}^{\infty}vvf(x,v;t)dv\label{eq:momflux}\end{equation}
Note that in one dimension the momentum flux is proportional to the
(kinetic) energy, but this is of course no longer true in higher dimension.

By multiplying both sides of the kinetic equation (\ref{eq:1dmodel})
successively by $1$, $v$ and $v^{2}$, and integrating over all
$v$, one easily arrives at the following macroscopic equations\begin{equation}
\partial_{t}n(x;t)+\partial_{x}J(x;t)=0\label{eq:mass-conserve}\end{equation}
\begin{equation}
\partial_{t}J(x;t)+\partial_{x}P(x;t)=-\gamma J(x;t)+n(x;t)a\label{eq:current-conserve}\end{equation}
\begin{equation}
\partial_{t}P(x;t)+\partial_{x}Q(x;t)=-2\gamma n(x;t)\frac{u^{2}(x;t)}{2}+2aJ(x;t)\label{eq:energy-conserve}\end{equation}
Eq.(\ref{eq:mass-conserve}) is the continuity equation expressing
the conservation of mass; equation (\ref{eq:current-conserve}) expresses
momentum balance with proper account of friction and acceleration
due to the electric field while (\ref{eq:energy-conserve}) is the
energy balance equation, taking into account the heat flux\begin{equation}
Q(x;t)=\int_{-\infty}^{\infty}v^{2}vf(x,v;t)dv\label{eq:heatflux}\end{equation}
as well as frictional dissipation. The standard kinetic equation with
the BGK collision operator yields the same three macroscopic equations
without the dissipative contributions stemming from the FP collision
operator. In the case of a steady, homogeneous flow, eq.(\ref{eq:current-conserve})
leads back to Ohm's law, $qJ=\sigma E$, with a conductivity $\sigma=nq^{2}/\gamma m$.

Stationary solutions of eq. (\ref{eq:1dmodel}) have been obtained
in the case of independent ions ($\omega=0$), for finite length as
well as infinitely long channels in ref. \cite{piasecki-allen-hansen}.
Solutions in the non-stationary case, and including the BGK collision
term in addition to the FP term, can only be obtained numerically.
In the next section we derive the lattice version of the kinetic equation
(\ref{eq:1dmodel}).

\section{The lattice-Fokker-Planck equation}

The lattice version of the kinetic equation (\ref{eq:1dmodel}) may
be systematically derived along the lines leading to the LB equation\cite{martys}.
Since the latter is well documented\cite{benzi-succi-vergassola,wolfe-gladrow,succi-book,chen-doolen}
we restrict most of the following discussion to the FP collision operator
(\ref{eq:1dFPoperator}).

The distribution function is expanded onto a Hermite basis\begin{equation}
f(x,v;t)=\sum_{k=0}^{K}F_{k}(x;t)h_{k}(v)w(v)\label{eq:hermite}\end{equation}
where $w(v)=(2\pi v_{T}^{2})^{-1/2}e^{-v^{2}/2v_{T}^{2}}$ is the
one-dimensional Hermite weight function, while $h_{k}(v)$ is the
Hermite polynomial of order $k$. By substituting eq.(\ref{eq:hermite})
into the kinetic equation eq.(\ref{eq:1dmodel}), and projecting upon
the Hermite basis, one arrives at \begin{equation}
\partial_{t}F_{l}(x;t)+\partial_{x}G_{l}(x;t)=C_{l}(x;t)\label{eq:FPlattice}\end{equation}
where\begin{eqnarray}
F_{l}(x;t) & = & \int_{-\infty}^{\infty}f(x,v;t)h_{l}(v)dv\\
G_{l}(x;t) & = & \int_{-\infty}^{\infty}vf(x,v;t)h_{l}(v)dv\\
C_{l}(x;t) & = & \int_{-\infty}^{\infty}C^{FP}[f(x,v;t)]h_{l}(v)dv\label{eq:cl}\end{eqnarray}
$C^{FP}$being the linear FP operator (\ref{eq:1dFPoperator}).

Equations (\ref{eq:FPlattice}) are the usual moment relations associated
with the FP equations. The next step is to evaluate the kinetic moments
$F_{l}$, $G_{l}$ and $C_{l}$ by Gauss-Hermite quadrature. Noting
that $f(x,v;t)/w(v)$ is a polynomial in $v$, the quadrature reads\begin{equation}
F_{l}(x;t)=\sum_{i=0}^{G-1}\frac{f(x,v_{i};t)}{w(v_{i})}w_{i}h_{l}(v_{i})\label{eq:quadrature}\end{equation}
where the $v_{i}$ and $w_{i}$ are the nodes and weights of the quadrature.
It is crucial to observe that Eq. (\ref{eq:quadrature}) is exact
for polynomials of degree up to $(2G+1)$. This means that the present
Hermite-Gauss projection is \emph{de-facto} equivalent to an optimized
form of importance sampling of velocity space. It is this discretization
which permits to compute the evolution of the low-order macroscopic
moments (density-current) more efficiently than any standard discretization
of velocity space (see ref. \cite{zhang-wei-kouri-hoffman} and references
therein). In fact, any grid method would necessarily use a mesh-spacing
in velocity space which is a fraction, say at most $1/10$, of the
thermal speed, $v_{T}$. Covering a few units of $v_{T}$ would then
take at least 30-50 grid points in velocity space. In contrast, LFPE
only needs at most five. The advantage to LFPE would be even more
substantial in higher dimensions. 

Expressions similar to Eq.(\ref{eq:quadrature}) hold for $G_{l}$
and $C_{l}$. Substituting these into eq. (\ref{eq:FPlattice}), and
identifying the factors of $h_{l}(v_{i})$ on both sides of the resulting
sum, one obtains the following set of equations\begin{equation}
\partial_{t}f_{i}(x;t)+v_{i}\partial_{x}f_{i}(x;t)=c_{i}(x;t)\,\,\,\,\,\,\,\,\,\,\,\,\,\,\,\,\,0\leq i\leq G-1\end{equation}
where the following identifications have been made\begin{equation}
f_{i}(x;t)\equiv\frac{f(x,v_{i};t)w_{i}}{w(v_{i})}\end{equation}
\begin{equation}
c_{i}(x;t)\equiv\frac{C^{FP}[f(x,v_{i};t)]w_{i}}{w(v_{i})}\label{eq:coeff}\end{equation}
The discrete collision operator is entirely specified by the coefficients
$c_{i}$ defined by eq. (\ref{eq:coeff}). These can be unambiguously
computed from the spectral decomposition of the continuous operator
$C^{FP}[f(x,v;t)]$ similar to eq. (\ref{eq:hermite}), i.e.\begin{equation}
C^{FP}(v_{i})\equiv C^{FP}[f(x,v_{i};t)]=\sum_{k=0}^{K}C_{k}(x;t)h_{k}(v_{i})w(v_{i})\end{equation}
Knowledge of the spectral coefficients $C_{k}(x,t)$ allows the discrete
coefficients $c_{i}(x;t)$ in eq.(\ref{eq:coeff}) to be calculated,
thereby providing an operational definition of the discrete FP operator.

The last step is to perform time integration along the characteristics
$dx_{i}=v_{i}dt$ according to the standard practice of LB algorithms.
This yields the desired lattice-Fokker-Planck (LFP) equation\begin{equation}
f_{i}(x+v_{i}\Delta t,t+\Delta t)-f_{i}(x,t)=c_{i}(x;t)\Delta t\label{eq:LFP}\end{equation}
where $\Delta t$ is the time-step chosen for the numerical solution.

The spectral coefficients of the FP operator are easily calculated
to be\begin{equation}
C_{0}=0\label{eq:c0}\end{equation}
\begin{equation}
C_{1}=-\gamma J+na\label{eq:c1}\end{equation}
\begin{equation}
C_{2}=-2\gamma(P-nv_{T}^{2})+2aJ\label{eq:c2}\end{equation}
\begin{equation}
C_{3}=-3\gamma(Q-3v_{T}^{2}J)+3a(P-nv_{T}^{2})\label{eq:c3}\end{equation}
\begin{equation}
C_{4}=-4\gamma(R-5v_{T}^{2}P+2nv_{T}^{4})+4a(Q-2Jv_{T}^{2})\label{eq:c4}\end{equation}
where $R=\int_{-\infty}^{\infty}dvv^{4}f(x,v;t)$ and the first four
Hermite polynomials are $h_{0}=1$, $h_{1}=v$, $h_{2}=v^{2}-v_{T}^{2}$,
$h_{3}=v^{3}-3vv_{T}^{2}$, $h_{4}=v^{4}-4v^{2}v_{T}^{2}+v_{T}^{4}$.
High order coefficients ($C_{k}$; $k\geq2$) differ from those derived
from the BGK operator (if $\omega=\gamma$). All coefficients, as
well as $n$, $u$, $P$ and $Q$, defined by eqs (\ref{eq:density})-(\ref{eq:momflux})
and (\ref{eq:heatflux}), and $R$, are local quantities, depending
on $x$ and $t$.

Returning to the LFP equation (\ref{eq:LFP}), we now consider two
discretized models, namely $Q_{3}$ and $Q_{5}$, involving three
and five velocities and where $v_{T}^{2}=1/3$ and $v_{T}^{2}=1$
respectively. Within $Q_{3}$ the velocities $v_{i}$ and associated
weights $w_{i}$ are\begin{eqnarray}
v_{0}=0 & \,\,\,\,\, & w_{0}=2/3\nonumber \\
v_{1}=+1; & \,\,\,\,\, & w_{1}=1/6\\
v_{2}=-1; & \,\,\,\,\, & w_{2}=1/6\nonumber \end{eqnarray}
The three velocities $v_{i}$ ($i=0,1,2$) may be considered as the
components of a three-vector $(0,1,-1)$. It proves convenient to
construct a basis of three such vectors ${\bf A}^{(k)}$ which satisfy
the orthonormality conditions\begin{equation}
{\bf A}^{(k)}\cdot{\bf A}^{(l)}=\sum_{i=0}^{2}A_{i}^{(k)}A_{i}^{(l)}w_{i}=\delta^{(k,l)}\end{equation}
These vectors are easily calculated to be\begin{eqnarray}
{\bf A}^{(0)} & = & (1,1,1)\nonumber \\
{\bf A}^{(1)} & = & \sqrt{3}(v_{0},v_{1},v_{2})=\sqrt{3}(0,1,-1)\\
{\bf A}^{(2)} & = & \frac{3}{\sqrt{2}}(v_{0}^{2}-v_{T}^{2},v_{1}^{2}-v_{T}^{2},v_{2}^{2}-v_{T}^{2})=\frac{1}{\sqrt{2}}(-1,2,2)\nonumber \end{eqnarray}
within this basis, the collision operator in eq. (\ref{eq:LFP}) may
be cast in the form\begin{equation}
c_{i}(x,t)=(C_{0}A_{i}^{(0)}+C_{1}A_{i}^{(1)}+C_{2}A_{i}^{(2)})w_{i}\end{equation}
The various $x$ and $t$-dependent quantities follow from the definitions
(\ref{eq:density})-(\ref{eq:momflux}) by replacing the integrals
over velocities by sums over the 3 discrete velocities $v_{0}$, $v_{1}$
and $v_{2}$, using the proper weights $w_{i}$. These prescriptions
entirely specify the Lattice Fokker-Planck (LFP) equation for the
model $Q_{3}$. The same procedure may be used to specify the LFP
equation for the $Q_{5}$ model, where the 5 velocities are $(-2,-1,0,1,2)$,
and the corresponding weights $w_{i}$ are $(1/12,1/6,1/2,1/6,1/12)$,
with a resulting thermal speed $v_{T}=1$.

The Lattice BGK equation has been widely described in the literature
\cite{succi-book}, and therefore only a very sketchy description
is provided below. The lattice BGK equation has the same form as eq.(\ref{eq:LFP}),
where the lattice BGK collision operator is given by\begin{equation}
c_{i}^{BGK}=-\omega(f_{i}-f_{BGK,i}^{eq})\label{eq:LBGK}\end{equation}
The local equilibrium takes the form of a local Maxwellian expanded
to second order in the Mach number $M=u/v_{T}$,\begin{equation}
f_{BGK,i}^{eq}=w_{i}n\left(1+\frac{v_{i}u}{v_{T}^{2}}+\frac{(v_{i}v_{i}-v_{T}^{2})u^{2}}{2v_{T}^{4}}\right)\label{eq:EQUIL}\end{equation}
The lattice equilibria, including the weights $w_{i}$, are designed
on the requirement of fulfilling mass and momentum conservation, that
is \begin{equation}
\sum_{i=0}^{G-1}f_{BGK,i}^{eq}=\sum_{i=0}^{G-1}f_{i}=n\label{eq:MASS}\end{equation}
and \begin{equation}
\sum_{i=0}^{G-1}f_{BGK,i}^{eq}v_{i}=\sum_{i=0}^{G-1}f_{i}v_{i}=nu\label{eq:MOM}\end{equation}
The parameter $\omega$ is an inverse relaxation scale which controls
the shear viscosity of the lattice fluid.

\section{Numerical results}

As a first application and test of the LFP formalism, we consider
the single-file transport of ions through a pore of length $L$ connecting
two reservoirs containing ions at given concentrations. As in ref.
\cite{piasecki-allen-hansen} we assume that at both ends of the pore,
the ion distribution function is a Maxwell-Boltzmann equilibrium distribution
at temperature $T$ which determines the thermal velocity $v_{T}$.
For instance, in the $Q_{3}$ case, boundary conditions are imposed
as follows\begin{eqnarray}
f_{1}(x=0,t) & = & \frac{n_{l}}{\sqrt{2\pi}}\nonumber \\
f_{2}(x=L+1,t) & = & \frac{n_{r}}{\sqrt{2\pi}}\label{eq:BCOND}\end{eqnarray}
where subscripts $1$ and $2$ stand for rightward and leftward propagation
and $n_{l}$, $n_{r}$ indicate the left and right reservoir densities
respectively. The reservoirs are located at $x=0$ and $x=L+1$ respectively,
while the physical channel runs from $1\le x\le L$. Note that the
expressions (\ref{eq:BCOND}) correspond to fixing the incoming \textit{fluxes}
from the reservoirs, hence they do not imply that $n_{l}$ and $n_{r}$
coincide with the fluid density at inlet and outlet sections. In fact,
since the ion distribution function in the reservoirs is a Maxwellian
at zero macroscopic speed, continuity of the fluxes (current density),
implies a discontinuity of both density and velocity profiles at both
ends of the channel. 

We define the dimensionless current\begin{equation}
J^{\star}=J\sqrt{\frac{2\pi}{v_{T}^{2}}}\frac{1}{n}\end{equation}
the dimensionless acceleration, or electric field\begin{equation}
a^{\star}=\frac{maL}{k_{B}T}=\frac{qEL}{k_{B}T}=\frac{E}{E_{T}}\end{equation}
and the dimensionless collision rate\begin{equation}
\gamma^{\star}=\frac{\gamma L}{v_{T}}=\frac{\gamma}{\gamma_{T}}=\frac{E_{\gamma}}{E_{T}}\end{equation}
where $n$ is the reservoir ion density, $E_{T}=k_{B}T/qL$ is a {}``thermal''
electric field, such that the work it produces to move a charge $q$
over the channel length $L$ equals the thermal energy $k_{B}T$,
while $E_{\gamma}=\gamma k_{B}T/(qv_{T})=m\gamma v_{T}/q$ is the
electric field producing on a charge $q$ a force which balances the
frictional force $-m\gamma v_{T}$. In other words, in a linear regime,
the drift velocities associated with $E_{T}$ and $E_{\gamma}$ are
$u_{T}=v_{T}/\gamma^{\star}$ and $u_{\gamma}=v_{T}$ respectively.
Clearly, in infinitely long channels, or zero-temperature fluids,
$u_{T}\rightarrow0$, and the ionic current is controlled by pure
dissipation, $u_{E}=\frac{qE}{m\gamma}$. In finite-size, finite-temperature
situations, and constant-flux boundary conditions, however, deviations
from this simple Ohmic regime must be expected, as we shall show in
the sequel.

We have solved the LFP equation numerically to determine the stationary
distribution function $f(x,v)$ and derive the corresponding macroscopic
moments, primarily the current $J(x)$. Since we focus on the Fokker-Planck
operator, for the time being, we exclude ion-ion collisions by setting
$\omega=0$.

In Figure 1, the stationary density profiles $n(x)$ calculated with
the $Q_{3}$ and $Q_{5}$ models for two values of the electric field,
$a^{\star}=1$ (moderate) and $a^{\star}=3$ (strong), are compared
to the analytical result of the continuous velocity model \cite{piasecki-allen-hansen},
which reads as follows: 

\begin{eqnarray*}
n_{exact}(x) & = & Ae^{a^{\star}x/L}+B\end{eqnarray*}
where \begin{eqnarray*}
A & = & \frac{1}{2\sinh(a^{\star}/2)}\left[n_{r}-n_{l}+\frac{a}{a^{\star}}\sqrt{2\pi}B\right]\\
B & = & \frac{n_{l}e^{a^{\star}/2}-n_{r}e^{-a^{\star}/2}}{2\sinh(a^{\star}/2)e^{-a^{\star2}/2\gamma^{\star2}}+\sqrt{\frac{\pi}{2}}\frac{a^{\star}}{\gamma^{\star}}\left[\cosh(a^{\star}/2)+\sinh(a^{\star}/2)\int_{-v_{T}a^{\star}/\gamma^{\star}}^{v_{T}a^{\star}/\gamma^{\star}}dv\Phi(v)\right]}\end{eqnarray*}
and $\Phi=\sqrt{1/2\pi v_{T}}\exp(-v^{2}/2v_{T}^{2})$. The total
current flowing through the channel is $J_{exact}=aB/\gamma$. It
should be appreciated that the strongly non-linear dependence of the
current (see the expression of $B$) reflects the highly non-trivial
competition between the effect of the boundary conditions and the
electric field.

The ratio of the numerical to analytical density profiles $n(x)/n_{exact}(x)$,
is seen to deviate from the unit value of about $10\%$ for $a^{\star}=3$
and less than $5\%$ for $a^{\star}=1$. These departures can be attributed
to the constant-flux boundary conditions which, as discussed previously,
introduce a discontinuity at the open ends of the channel. We also
notice that the $Q_{5}$ solution appears appreciably more accurate
than the $Q_{3}$ one. At lower values of the electric field, both
$Q_{3}$ and $Q_{5}$ yield excellent agreement with the analytical
solution. 

The current-voltage relation is illustrated in Fig. 2, where the reduced
current $J^{\star}$ is plotted as a function of the reduced electric
field, $a^{\star}\equiv E/E_{T}$, for three values of the reduced
collision rate $\gamma^{\star}=E_{\gamma}/E_{T}=1,5,10$. Results
obtained with the $Q_{3}$ and $Q_{5}$ models are compared to the
analytical predictions of ref. \cite{piasecki-allen-hansen}, where
the problem was solved for a continuous velocity spectrum ($Q_{\infty}$).
First, we notice that the Ohmic regime, $E/E_{T}\ll1$, is well reproduced
in all three cases. As the strength of the electric field is increased,
the current shows the saturation effect predicted by the analytical
results \cite{piasecki-allen-hansen}.

This saturation is a direct consequence of the imposed boundary conditions.
Since the reservoirs cannot feed more than $nv_{T}$ ions per unit
area and time, the current cannot exceed its ballistic value $v_{T}n/\sqrt{2\pi}$
in the limit $E\rightarrow\infty$. The rate of convergence towards
the asymptotic limit depends sensitively on the reduced collision
rate $\gamma^{\star}$. Fig. 2 shows that the results obtained with
the $Q_{5}$ model are systematically closer to the exact results
than those obtained with the $Q_{3}$ model.

We have also checked that with simple boundary conditions, e.g. periodic,
Ohm's law $J^{\star}=E^{\star}/\gamma^{\star}$, is reproduced to
machine accuracy and independently on the grid size, beyond about
$100$ grid points. However, the present case is much more challenging
due to the highly non-trivial boundary conditions. In fact, a very
slow convergence of the results as a function of grid resolution is
observed (see Figure 3). This is probably caused by the representation
of the reservoirs by a single lattice point without making allowance
for any finite size transition region between the current-carrying
distribution function in the channel and the no-current equilibrium
current in the reservoirs. The consequence of a less abrupt treatment
of the boundaries will be explored in the future.

\section{Effects of ion-ion collisions}

The BGK collision operator (\ref{eq:LBGK}) does not make any contribution
to the continuity and momentum equations because, by construction,
ion-ion collisions have been designed to conserve mass and momentum
(elastic collisions). As a result, the BGK operator does not contribute
to the momentum equation, and consequently, it cannot produce any
effect on the electric conductivity of the system. In fact, ion-ion
collisions are in control of the shear viscosity of the charged fluid,
clearly an irrelevant notion in one-dimensional systems. 

However, in electro-rheological flows, ions move in the presence of
a surrounding solvent, typically water. It is therefore reasonable
to assume that ion-ion {}``effective'' collisions may become inelastic
on account of the ion interaction with the solvent. A natural way
to include such inelastic effects within the BGK operator is to assume
a mismatch between the equilbrium current $J^{eq}=\sum_{i=0}^{G-1}f_{i}^{eq}v_{i}$
and the actual current $J$ carried by the ions. The simplest form
for such a mismatch is \begin{equation}
J^{eq}=\lambda J\label{eq:inelastic}\end{equation}
where $\lambda$ is a parameter measuring the degree of inelasticity
of the 'effective' ion-ion collisions ($0<\lambda<1$ for passive
solvents, i.e. momentum absorbing media, while $\lambda>1$ denotes
active solvents, i.e. momentum-imparting media, and $\lambda=1$ for
standard elastic collisions). 

The relation (\ref{eq:inelastic}) is readily implemented by replacing
$u$ with $\lambda u$ in the second term within parentheses on the
right hand side of the eq. (\ref{eq:EQUIL}). With this modification,
the BGK collision operator contributes a term $-\omega(J-J^{eq})=-\omega(1-\lambda)J$
to the momentum equation, so that inelastic collisions make themselves
felt through an effective frequency $\omega_{\lambda}=(1-\lambda)\omega$.
As a result, setting aside boundary conditions, i.e. in an infinitely
long channel, for simplicity, the steady-state current is given by\[
J=\frac{nE/\gamma}{1+\omega_{\lambda}/\gamma}\]
This expression has been checked against numerical simulations (with
periodic boundary conditions) for $\lambda=0$ (fully inelastic) and
$\lambda=1$ (fully elastic), with $\gamma=0.1$ and $\omega/\gamma=0.1,1,10$.
In Figure 4, we report the ratio $J(\lambda=0)/J(\lambda=1)$ as a
function of the electric field strength, and compare it with the analytical
result $\gamma/(\gamma+\omega)$. From this figure, excellent agreement
with the analytical results is clearly appreciated. Although the present
test only serves the purpose of illustrating the basic idea, it is
hoped that combination of the lattice Fokker-Planck equation with
inelastic BGK operators, will permit to explore complex situations
in one, two and three dimensions, out of reach of analytical methods,
such as multicomponent fluids (ions, cations and solvent) with heterogeneous
surface interactions, trapping effects and related phenomena.

\section{Conclusion}

In conclusion, we have developed a lattice version of the Fokker-Planck
equation (FPE) which, by construction, can solve near-equilibrium
kinetic problems in the full single-particle phase space. The main
scope of the lattice FPE is to account for dissipative interactions
not resolved at the molecular scale, such as fluid interactions with
solid walls and/or solute-solvent collisions. The lattice FPE has
been applied to the study of electrorheological transport of a one-dimensional
charged fluid, and found to yield satisfactory agreement with a recent
non-trivial analytical solution, especially for the five-speed case.
In particular, the lattice FPE proves capable of predicting the saturation
effect resulting from the non-linear interaction between the electric
field and the constant-flux boundary conditions imposed by the presence
of equilibrium reservoirs at the channel boundaries. 

The present lattice FPE extends straightforwardly to higher dimensions
and it might prove useful for the numerical investigation of more
complex situations, such as heterogeneus channels with kinetic traps,
and/or multicomponent fluids, for which analytical solutions are no
longer available. 

\begin{acknowledgments}
JPH acknowledges the generous support of INFM while in Rome and the
kind hospitality of G. Ciccotti and G. Parisi. The authors are grateful
to D.Moroni, B. Rotenberg and A. Louis for useful comments and discussions.
\end{acknowledgments}
\appendix

\section{Stability analysis of the Lattice FPE }

Being derived from the same principles, the lattice Fokker-Planck
equation shares many features with the Lattice Boltzmann equation.
Among the main advantages: i) a very efficient sampling of velocity
space, which permits to work in steps of size $\Delta v=v_{T}$ rather
than of a fraction thereof; ii) since the discrete speeds $v_{i}$
are constant, the streaming operator $v\partial_{x}f$ can be integrated
\emph{exactly} along the characteristics $\Delta x=v_{i}\Delta t$,
i.e. by a mere shift of the discrete distribution from site $x$ to
site $x+v_{i}\Delta t$, iii) the collision operator is completely
local in space, which makes the Lattice FPE well suited to parallel
computing. Of course, there are limitations too. In particular, the
damping rate is approximately bounded within the range \[
a/v_{T}<\gamma<1/\Delta t\]
 The lower bound relates to the fact that damping rates below $a/v_{T}$
imply that particles may acquire a drift speed $u=a/\gamma$ larger
than the thermal speed. This endangers the stability of the lattice
FPE, because the condition $u>v_{T}$ may violate the positive-definiteness
of the truncated distribution (eq.\ref{eq:hermite}), which consists
only of a very limited number of Hermite polynomials. Thus, in full
analogy with LBE, the lattice FPE is best suited to describe near-equilibrium,
low-Mach number flows, with $u/v_{T}\ll1$. The upper boundary is
dictated by a stability condition of the time-marching scheme, specifically
of the explicit time integration of the collision operator (we recall
that the streaming operator is integrated exactly). The stability
of the lattice FPE can be estimated by means of the following inequality
\begin{equation}
|\frac{C_{k}\Delta t}{F_{k}}|<1\label{STAB}\end{equation}
 which states that the change in the macroscopic moment $F_{k}(x,t)$
due to collisions in a time step $\Delta t$, should not exceed the
value of $F_{k}(x,t)$ itself.

In order to elaborate this condition further, we recall the expression
of the five Hermite polynomials relevant to the $Q_{5}$ lattice,
that is: \[
h_{0}(v_{i})=1_{i}=[1,1,1,1,1]\]
 \[
h_{1}(v_{i})=v_{i}=[-2,-1,0,+1,+2]\]
 \[
h_{2}(v_{i})=v_{i}^{2}-v_{T}^{2}=[3,0,-1,0,3]\]
 \[
h_{3}(v_{i})=v_{i}^{3}-3v_{T}^{2}v_{i}=[-2,2,0,-2,2]\]
 \[
h_{4}(v_{i})=v_{i}^{4}-4v_{T}^{2}v_{i}^{2}+v_{T}^{4}=[1,-2,1,-2,1]\]
 The corresponding macroscopic moments in equations (\ref{eq:c0})-(\ref{eq:c4}),
are given by: \begin{eqnarray}
F_{0} & = & n\label{HERMITE}\\
F_{1} & = & J\\
F_{2} & = & nu^{2}\\
F_{3} & = & Q-3Jv_{T}^{2}\\
F_{4} & = & R-4nv_{T}^{2}u^{2}-3nv_{T}^{4}\end{eqnarray}
 In view of these expressions, the stability condition (\ref{STAB})
is readily recast in a more informative form as: \begin{equation}
k\gamma\Delta t<1,\; k=1,4\label{STAB1}\end{equation}
 A more specific result can be obtained by analyzing the dispersion
relation associated with the lattice FPE. Upon Fourier-transforming
the lattice FPE eq.(\ref{eq:LFP}), $f_{i}(x,t)=\sum_{k,\omega}f_{i}(k,\omega)e^{I(kx-\omega t)}$,
we obtain ($I$ denotes the imaginary unit): \begin{equation}
\sum_{j}[(e^{-I(\omega-kv_{i})\Delta t}-1)\delta_{ij}-C_{ij}\Delta t]f_{j}=0\label{DR}\end{equation}
 where $C_{ij}$ is the collision matrix associated with the Fokker-Planck
operator. This matrix can be computed as follows. Consider the definition
of the spectral coefficient $C_{l}$ in eq.(\ref{eq:cl}), and express
the function $C^{FP}[f]$ as $\hat{C}f$, where $\hat{C}=\partial_{v_{\alpha}}[R_{\alpha}+D\partial_{v_{\alpha}}]$
is the Fokker-Planck operator acting upon $f(x,v,t)$. By expanding
the function $f$ on the Hermite basis as given by eq.(\ref{eq:hermite}),
eq.(\ref{eq:cl}) takes the form \begin{equation}
C_{k}=\sum_{l}C_{kl}F_{l}\end{equation}
 where \begin{equation}
C_{kl}=\int h_{k}(v)\hat{C}\, w(v)h_{l}(v)dv\label{CLM}\end{equation}
 is the matrix representation of the Fokker-Planck operator in the
global Hermite basis. By using eq.(\ref{eq:quadrature}) to express
$F_{l}$ in terms of $f_{j}$, we obtain: \[
C_{i}=\sum_{kl}C_{kl}h_{k}(v_{i})w_{i}\sum_{j}h_{l}(v_{j})f_{j}\equiv\sum_{j}C_{ij}f_{j}\]
 which defines the collision matrix $C_{ij}$ as: \begin{equation}
C_{ij}=w_{i}\sum_{kl}h_{k}(v_{i})C_{kl}h_{l}(v_{j})\label{CIJ}\end{equation}
 The above expression is the operational key of the lattice FPE and
lends itself to a fairly transparent physical interpretation. The
matrix element $C_{ij}$, expressing the effect of population $f_{i}$
on population $f_{j}$, is a weighted average over all spectral modes,
the weights being the Hermite eigenfunctions evaluated at $v=v_{i}$
and $v=v_{j}$ respectively.

It is now convenient to normalize the Hermite coefficients $h_{k}(v_{i})$
as follows: \[
h_{ik}\equiv h_{k}(v_{i})/\sqrt{H_{k}}\]
 where $H_{k}=\sum_{i}h_{k}(v_{i})w_{i}h_{k}(v_{i})$ are the normalization
factors, explicitly $H_{0}=1,H_{1}=1,H_{2}=2,H_{3}=2,H_{4}=2$.

The expressions (\ref{CIJ}) and (\ref{CLM}) are nothing but the
matrix representations of the Fokker-Planck operator in the local
(Dirac's deltas) and global (Hermite polynomials) basis functions.
The transformation between these two representations is performed
by the matrix $h_{ik}$. It is readily checked that this matrix fulfills
the orthonormality condition \[
\sum_{k}h_{ik}w_{i}h_{kj}=\delta_{ij}\]
 Consequently, the matrices $C_{ij}$ and $C_{kl}$ are related by
a similarity transformation, hence they are iso-spectral (they share
the same eigenvalues).

The spectrum of $C_{kl}$ can be obtained by direct inspection of
eq.s (\ref{eq:c0})-(\ref{eq:c4}) for the case $E=0$, and taking
into account the definitions (\ref{HERMITE}). The resulting matrix
$C_{kl}$ is identified as: \[
C_{kl}=\left(\begin{array}{ccccc}
0 & 0 & 0 & 0 & 0\\
0 & -\gamma & 0 & 0 & 0\\
0 & 0 & -2\gamma & 0 & 0\\
0 & 0 & 0 & -3\gamma & 0\\
0 & 0 & 4\gamma & 0 & -4\gamma\end{array}\right)\]
 which delivers the following five eigenvalues: \[
\lambda_{k}=-k\gamma,\;\;\;\;\;\; k=0,4\]
 Note the zero eigenvalue associated with mass conservation.

According to standard arguments of Lattice Boltzmann theory \cite{benzi-succi-vergassola},
the stability condition associated with the dispersion relation (\ref{DR}),
reads as follows: \begin{equation}
|1-k\gamma\Delta t|<1,\;\;\;\; k=1,4\end{equation}
 This is satisfied within the range \begin{equation}
0<\gamma\Delta t<1/2\end{equation}

It is perhaps interesting to observe that the procedure outlined in
this Appendix can be applied to a very broad class of kinetic equations,
including the Klein-Gordon and Schr\"odinger equations of quantum
mechanics \cite{succibenziquantum}.

\bibliographystyle{prsty}
\bibliography{/home/smelch/1.TESTI/FOKKERPLANCK/biblio}

\includegraphics[%
  scale=0.8,
  angle=-90]{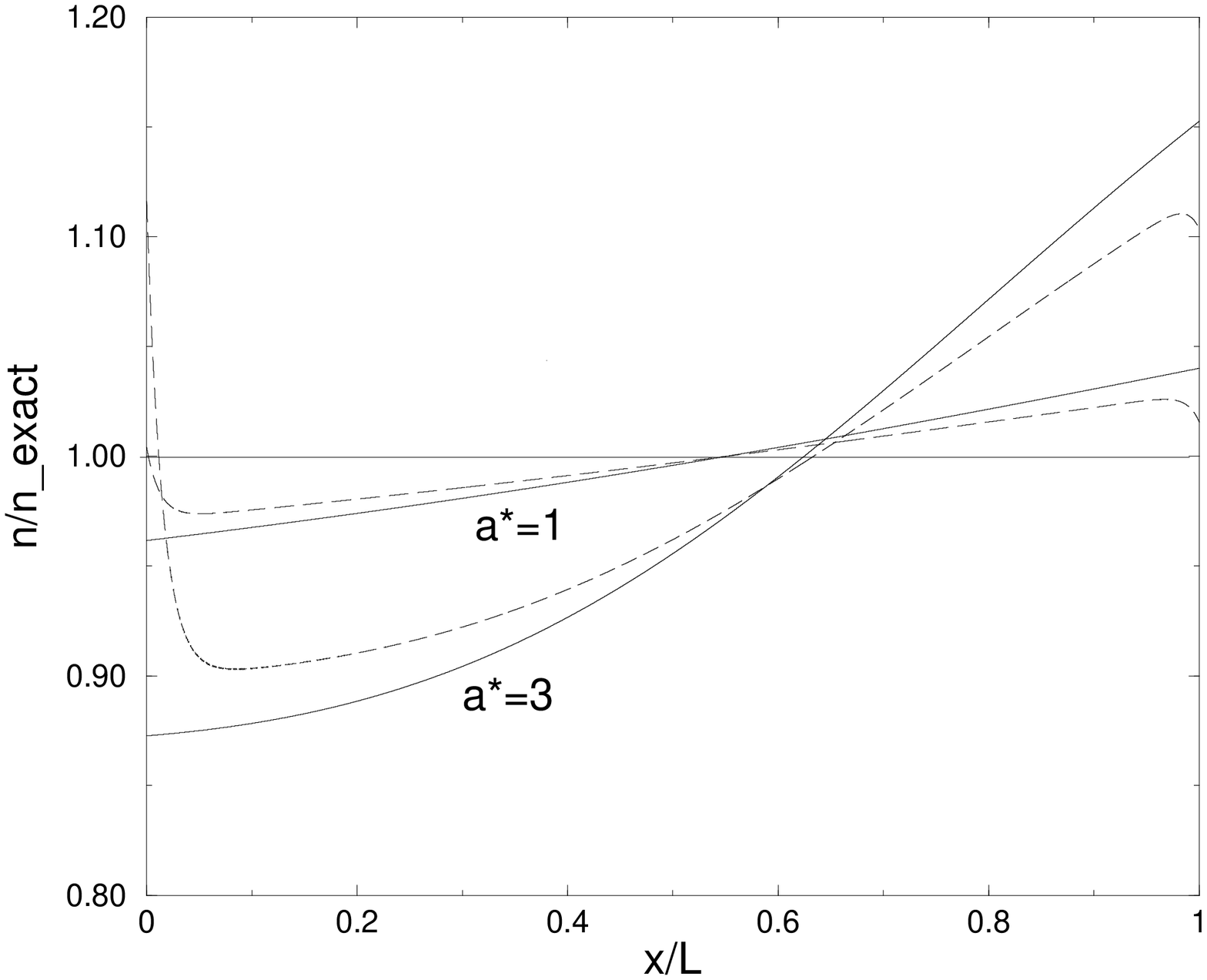}

\begin{figure}

\caption{Ratio of numerical to analytical density profiles along a channel
of length $10000$, for $a^{\star}=1.0$ and $a^{\star}=3.0$. The
solid line and the dashed line correspond to the $Q_{3}$ and the
$Q_{5}$ models respectively.}
\end{figure}

\includegraphics[%
  scale=0.8,
  angle=-90]{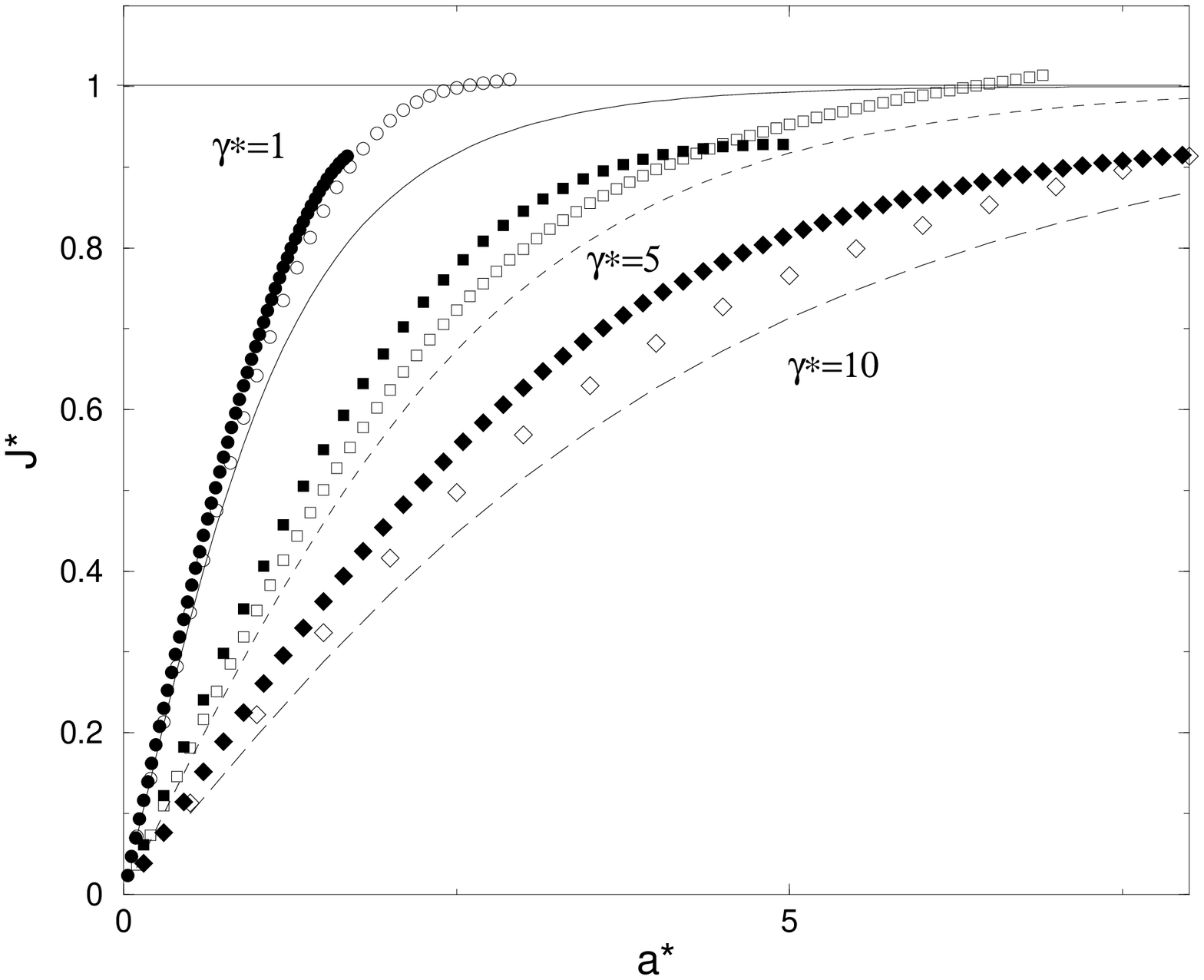}

\begin{figure}

\caption{Reduced current $J^{\star}$ as a function of the reduced applied
field $a^{\star}=E/E_{T}$. Filled symbols correspond to the $Q_{3}$
model while open symbols to the $Q_{5}$ model. Circles, squares and
diamonds correspond to $\gamma^{\star}=0.1$, $5$ and $10$ respectively.
Solid, dashed and long-dashed lines are the theoretical predictions
for the three values of $\gamma^{\star}$ . The horizontal line highlights
the limiting plateau $J^{\star}=1$.}
\end{figure}

\includegraphics[%
  scale=0.8,
  angle=-90]{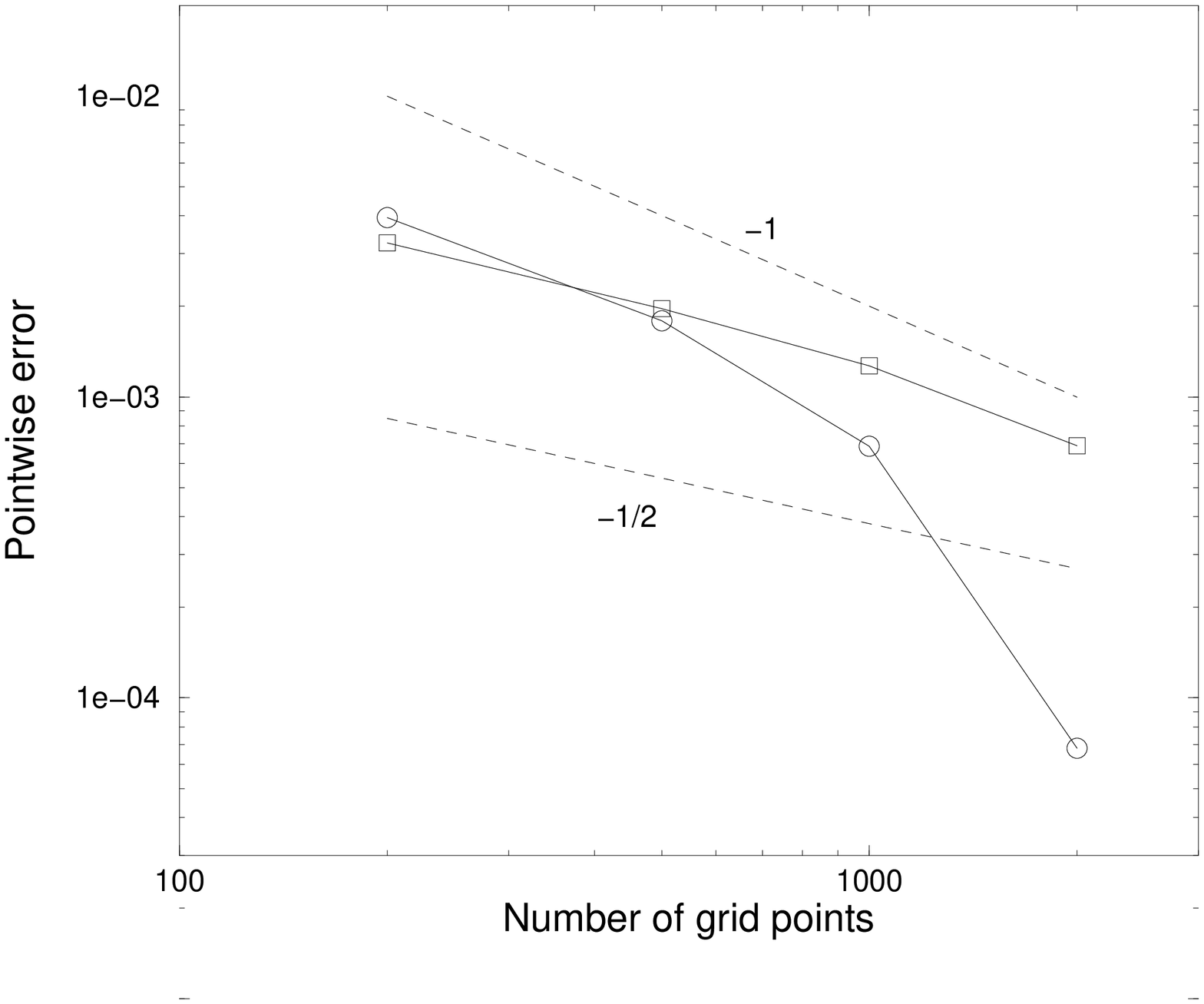}

\begin{figure}

\caption{The pointwise error of the numerical vs. analytical solution for
the density and current at the midpoint of the channel for the $Q_{5}$
lattice as a function of the number of grid points $N$. Circles and
squares refer to density and current respectively. The main parameters
are $a^{\star}=1$ and $\gamma^{\star}=10$. The dashed lines correpond
to $N^{-1}$ and $N^{-1/2}$ convergence.}
\end{figure}

\includegraphics[%
  scale=0.8,
  angle=-90]{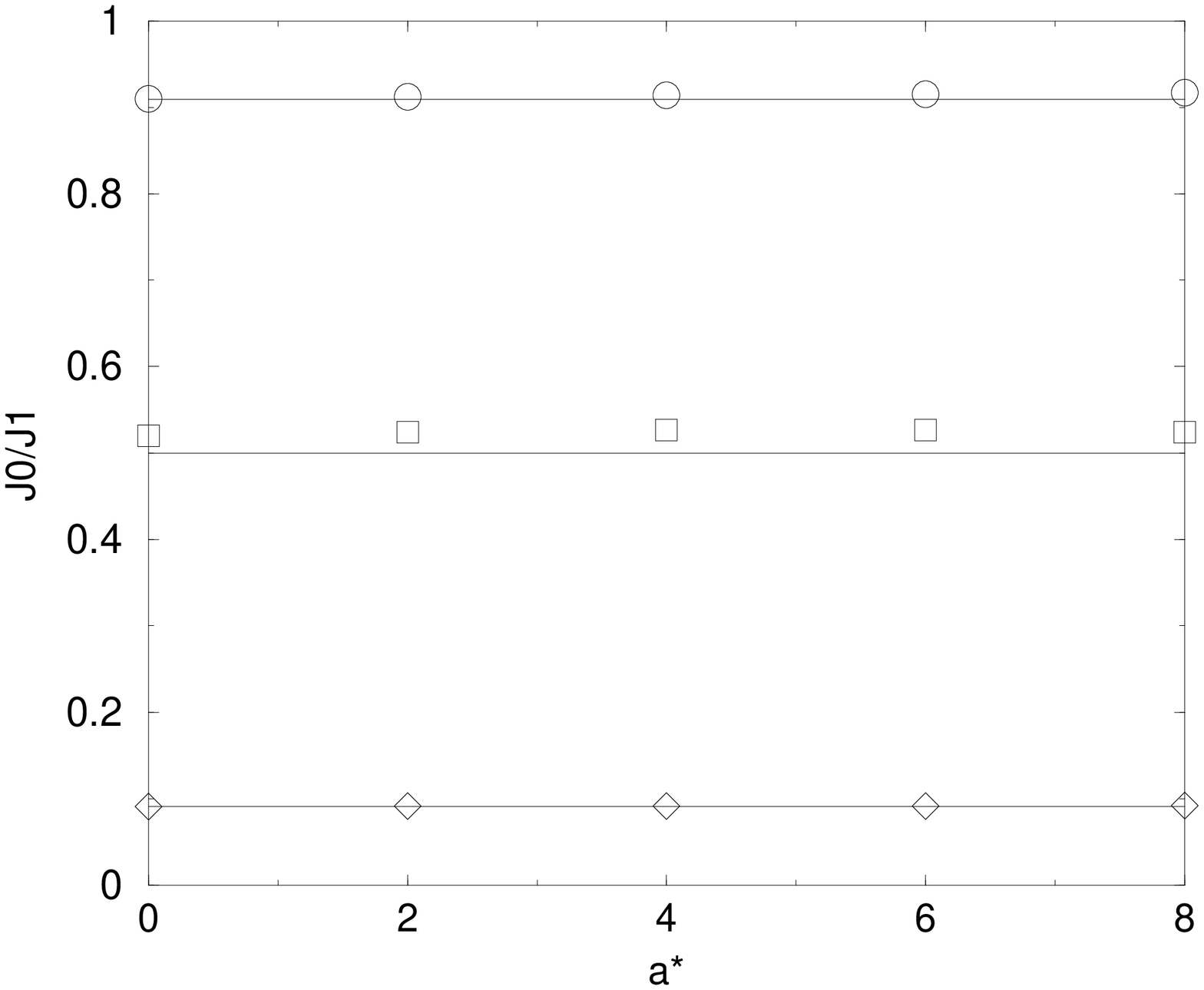}

\begin{figure}

\caption{Ratio of currents for elastic vs. inelastic collisions, $J_{\lambda=0}/J_{\lambda=1}$,
for $E_{\gamma}/E_{T}=100$. Three values of $\lambda/\omega=0.1$,
$1$, and $10$ correspond to the upper, middle and lower curves respectively.
The horizontal lines represent the theoretical prediction.}
\end{figure}

\end{document}